\documentclass[12pt]{elsarticle}
\usepackage{graphicx} % Required for inserting images
\usepackage{array}
\usepackage{tabularx}
\usepackage{hyperref}
\usepackage{amsfonts} 
\usepackage{breqn}
\usepackage{multirow}
\usepackage{pdflscape}
\usepackage{float}
\usepackage{caption}
\usepackage{mathrsfs}
\usepackage{amsmath}
\usepackage{setspace}
\usepackage{breqn}
\usepackage[labelfont=bf,labelsep=period]{caption}
\usepackage[a4paper, left=2.5cm, right=2.5cm, top=3.5cm, bottom=2.5cm]{geometry}
\usepackage{etoolbox}

\biboptions{sort&compress}

\hypersetup{
 colorlinks,
 citecolor=black,
 filecolor=black,
 linkcolor=black,
 urlcolor=black
}

\title{\textbf{Fast and Accurate Inverse Blood Flow Modeling from Minimal Cuff-Pressure Data via PINNs}}
\author[inst1]{Sokratis J. Anagnostopoulos}
\author[inst1]{Georgios Rovas}
\author[inst1]{\\Lydia Aslanidou}
\author[inst3]{Vasiliki Bikia}
\author[inst1]{Nikolaos Stergiopulos}

\affiliation[inst1]{organization={Laboratory of Hemodynamics and Cardiovascular Technology, EPFL},
   city={Lausanne},
   country={Switzerland}}
   
\affiliation[inst3]{organization={Stanford DBDS and HAI, Stanford University},
   city={Palo Alto},
   country={USA}}
   
\begin{document}
\begin{abstract}

Accurate assessment of central hemodynamics is essential for diagnosis and risk stratification, yet it still relies largely on invasive measurements or on indirect reconstructions built from population-averaged transfer functions. While conventional methods are valuable in clinical practice, they face limitations, particularly in personalized medicine. Physics-informed methods address these by integrating physical principles, reducing the need for extensive data. In this work, a fully noninvasive, patient-specific framework is developed that combines a validated 1-D model of the systemic arterial tree with physics-informed neural networks (PINNs). This model performs the inverse solution of the flow and pressure fields within the arterial network, given minimal noninvasive measurements of pressure from a cuff reading and trains in 4000 iterations, at least 10x faster than the current state-of-the-art models due to several model enhancements. We validate the model predictions against our 1-D solver, yielding a near perfect correlation, and perform additional tests on a clinical dataset for the identification of important central hemodynamic parameters of cardiac output $CO$ and central systolic blood pressure $cSBP$, with correlations of $r=0.847$ and $r=0.951$, respectively. Moreover, the model is able to tune the patient-specific coefficients of the terminal resistance $R_T$ and compliance $C_T$ while training, treating them as learnable parameters. The inverse PINN model is able to solve the entire tree of 8 arteries with a single network, costing 5-10 minutes of computational time. This significant performance boost compared to traditional iterative inverse methods holds promise towards applications of personalized cardiac output monitoring and hemodynamic assessment via noninvasive approaches like wearable devices.

\end{abstract}
\maketitle

\newpage

\section{Introduction}
The field of cardiovascular mechanics combines many different aspects of complex physics interactions: from non-Newtonian pulsating flow through an intricate system of vascular branching, to the non-linear visco-elasticity of the arterial wall, and the coupling between them. LHTC has focused on the 1-D modeling of the arterial network developing a generic 1-D model \cite{reymond2009validation} and validating it against noninvasive measurements, offering valuable insights into flow and pressure waveforms across vascular beds. The numerical model encompasses systemic, coronary and cerebral arteries, along with heart dynamics, and solves the Navier-Stokes equations incorporating a time-varying elastance heart model and considering systolic flow impediment in coronary arteries. It has been extensively applied on cardiovascular physiology, exploring the effects of arterial stiffening and aortic aneurysm reconstruction on wave reflections and central blood pressure \cite{reymond2012systolic,vardoulis2011impact}, and to provide patient-specific \cite{reymond2011validation,bikia2019noninvasive} solutions. Moreover, it has been combined with advanced computational techniques like 3-D CFD and FSI to understand complex flow phenomena in the aorta and cerebral vasculature \cite{augsburger2011intracranial}, enabling the development and validation of noninvasive techniques for assessing vascular health \cite{papaioannou2012systolic,vardoulis2012estimation,papaioannou2014vivo}.

Advances in machine learning (ML) are paving the way towards exploiting the information encapsulated in noninvasive, noisy biomedical measurements ranging from peripheral pressure data to Magnetic Resonance Imaging (MRI). These models are deployed in order to make fast and accurate predictions of important physiological indicators of the patient’s cardiovascular health like the central arterial pressure \cite{xiao2022reconstruction}, cardiac contractility \cite{bikia2020noninvasive}, or the Wall Shear Stress field (WSS) \cite{su2020generating}. A recent breakthrough in ML for physics were physics-informed neural networks (PINNs), which are not trained on just raw data but also on the actual differential equations that govern the system (e.g. Navier-Stokes) \cite{raissi2019physics,lu2021deepxde}. These networks have been previously applied on cardiovascular systems, assimilating noisy MRI data and identifying hemodynamic parameters regarding the patient-specific characteristics \cite{kissas2020machine,sarabian2022physics,garay2024physics}, or to recreate the full velocity, transient cuffless pressure or shear stress field from sparse measurements \cite{arzani2021uncovering,sel2023physics}. PINNs have gained increased popularity due to their superiority in inverse problems, motivating researchers to develop methodologies to ease the optimization process. However, the state-of-the art models often require many hours to train for a single patient geometry, which does not qualify them as efficient methods in clinical practice yet.

Therefore, the aim of this study is to investigate the application of an efficient PINN framework for the rapid solution of the PDEs governing the cardiovascular system, leveraging the validated arterial model developed at Laboratory of Hemodynamics and Cardiovascular Technology (LHTC). The proposed model enables the use of a single neural network to simultaneously compute the velocity, pressure, and area fields across various arterial segments. We demonstrate that accurate predictions of flow and pressure pulses within arterial bifurcations can be inferred using a single noninvasive pressure measurement at the brachial artery. With the integration of PINN enhancements, the methodology provides patient-specific results within 5-10 minutes. In contrast, traditional inverse solution methods, which rely on multiple forward solutions with our 1-D numerical model, typically require several hours to compute. Similar costs are required by other physics-informed approaches currently available in the literature.

\section{Methods}

\subsection{1-D arterial model}

Our 1-D cardiovascular model treats arteries as tapered segments with viscoelastic flexible walls. The 1-D continuity and momentum equations (Navier-Stokes) are solved iteratively via an implicit scheme for an arterial network consisting of 103 conical segments, along with a constitutive law for the flexible arterial walls \cite{reymond2009validation}, given by the following closed-system of PDEs:

\begin{equation}
\frac{\partial Q}{\partial t} + \frac{\partial}{\partial x} \left( \int_A u^2dA\right) = -\frac{A}{\rho}\frac{\partial P}{\partial x} - 2\pi R \frac{\mu}{\rho} \frac{\partial u}{\partial r} \Bigg|_{r=R} + A f_x,
\label{p4:eq1}
\end{equation}
\begin{equation}
\frac{\partial P}{\partial t} = -\frac{1}{C_A}\left(\frac{\partial A^v}{\partial t} + \frac{\partial Q}{\partial x}\right),
\label{p4:eq2}
\end{equation}
\begin{equation}
\frac{\partial A}{\partial P} = \frac{A_{ref}}{\rho \cdot PWV^2} \cdot \left[ a + \frac{b}{1+\left( \frac{P-P_{maxC}}{P_{width}}\right)^2}\right]= C_A,
\label{p4:const}
\end{equation}

\noindent where $A(x, t)$ is the arterial lumen area of radius $R(x, t)$, $Q(x, t)$ is the volumetric blood flow rate, $P(x, t)$ is the blood pressure, $\mu$ is the blood viscocity, $f_x$ is the gravitational force, $P_{maxC}, P_{width}$ are constants, and $u(x, r, t)$ is the velocity profile approximated by the Witzig-Womersley theory. The constitutive law (Eq. \ref{p4:const}), depends on the area compliance $C_A$, which is a function of arterial pulse wave velocity ($PWV$) and pressure, with $A_{ref}$ being the reference cross-sectional area at $P_{ref}=100$ mmHg. The conservation of mass and pressure is applied at the interface of each arterial bifurcation, where a parent vessel $p$ (with axial coordinate $x = L_p$ at the downstream end) splits into $N_d$ daughter vessels $d_i$ (with coordinates $x = 0$ at their upstream ends). These additional constrains are given by:
\begin{equation}
Q_p(L_p,t)
=
\sum_{i=1}^{N_d} Q_{d_i}(0,t),
\label{p4:eq:mass_junction}
\end{equation}
\begin{equation}
P_p(L_p,t)
=
P_{d_1}(0,t)
=
P_{d_2}(0,t)
=
\cdots
=
P_{d_{N_d}}(0,t).
\label{p4:eq:pressure_junction}
\end{equation}

Finally, at the terminal arterial sites, a 3-element Windkessel model is used to simulate the omitted branches with their corresponding terminal resistance $R_T = R_1 + R_2$ and compliance $C_T$ parameters, which can be fine-tuned based on the patient, hence:

\begin{equation}
\frac{\partial Q}{\partial t} = \frac{1}{R_1}\frac{\partial P}{\partial t} + \frac{P}{R_1R_2C_T} + \left(1+ \frac{R_1}{R_2}\right)\frac{Q}{R_1C_T},
\label{p4:wk3}
\end{equation}

\subsection{Physics-Informed Neural Network (PINN) model}

A PINN ($f_{\theta}$) approximates the solution of a partial differential equation (PDE) given by:
\begin{equation}
\mathcal{D}\{f(x,t)\} = q(x,t),
\end{equation}
\noindent where \(f(x,t)\) is the unknown function we wish to approximate, \(\mathcal{D}\) denotes the differential operator, and \(q(x,t)\) is a given source or forcing function that introduces external influences to the system. The differential operator \(\mathcal{D}\) depends on the specific PDE under consideration. The network primarily aims to minimize the residuals $\mathcal{R} = \mathcal{D}\{f(x,t)\} - q(x,t)$, along with any observed data. Additionally, the problem may be constrained by several types of boundary conditions (e.g.\ conservation at bifurcations and Windkessel conditions at terminal arteries).

\begin{figure}[H]
 \centering
 \includegraphics[width=1.0\textwidth]{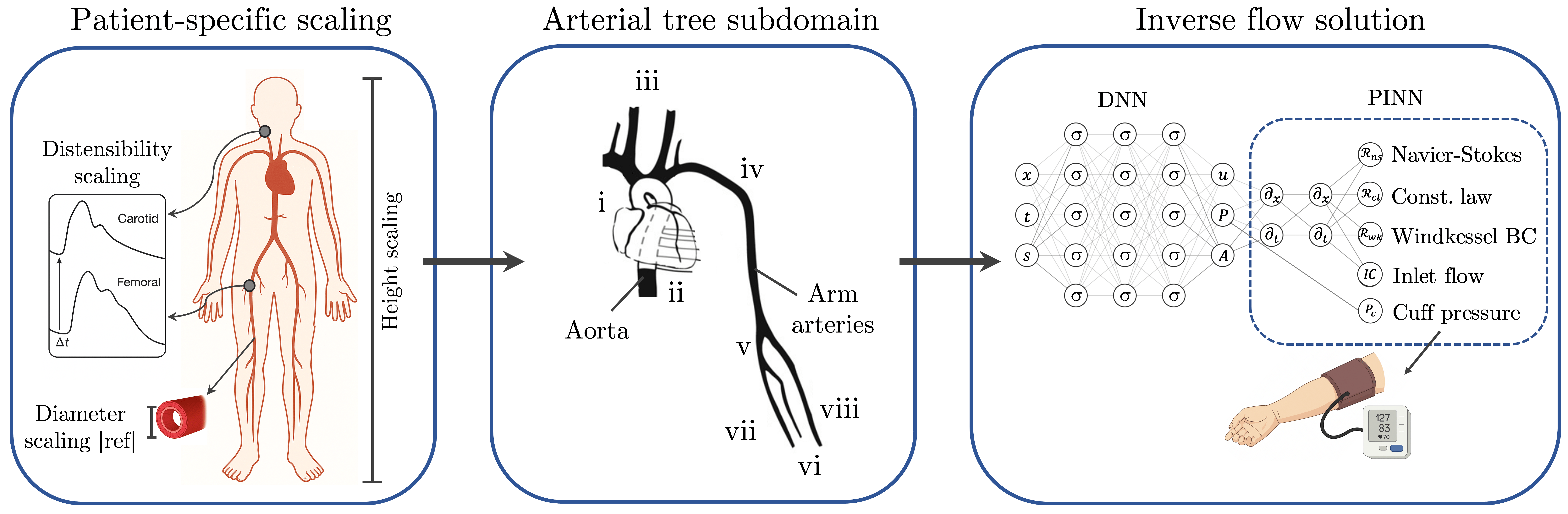}
 \caption[Inverse cardiovascular flow framework]{ \textbf{Inverse solution framework:} The 1-D arterial network is adjusted on a patient-specific basis, so that the geometry and arterial compliance are matched based on the age, height, gender and $cfPWV$. Then, an arterial subdomain is extracted from the reference tree with adjusted terminal resistance/compliance parameters. Finally, an inverse flow solution is performed by training the PINN model to match the cuff pressure measurements, while adjusting/learning the patient-specific $CO$ and Windkessel parameters.}
 \label{p4:workflow}
\end{figure}

The loss function $\mathcal{L}$ in PINNs is what drives the training process to the optimal state and is designed to encompass the deviation of the neural network prediction from the PDE solution and also the available measurements:
\begin{equation}
\mathcal{L} = \mathcal{L}_{IC} + \mathcal{L}_{BC} + \mathcal{L}_{\mathcal{R}} + \mathcal{L}_{data} ,
\label{p4:eq:loss}
\end{equation}

\noindent where \(\mathcal{L}_{IC}\), \(\mathcal{L}_{BC}\) represent the residuals associated with the initial and boundary conditions of the problem, \(\mathcal{L}_{\mathcal{R}}\) are the PDE residuals (calculated during training with automatic differentiation) and \(\mathcal{L}_{data}\) is the deviation from the measured data. It is important to note that a conventional neural network would need a large dataset, as it is only trained using the term \(\mathcal{L}_{data}\). On the other hand, PINNs exploit the known physics through the additional loss terms and require minimal measurements for the training process, which makes them ideal in a clinical setting where the available data may be scarce or noisy. Hence, they are designed to provide a seamless integration of the governing physical laws with the case-specific experimental data, representing a fast inverse solver. The PINN framework is shown in Fig. \ref{p4:workflow}, where we start by scaling a reference geometry \cite{reymond2009validation}, based on the patient specific characteristics. For the scaling we adopt experimental correlations of the literature \cite{wolak2008aortic}, and apply global multipliers for the length, diameter and compliance. A truncated arterial subdomain is then considered for the inverse solution via the PINN framework, which seamlessly integrates the peripheral blood flow characteristics (e.g. cuff measurements for $SBP$ and $DBP$) with central hemodynamics (i.e. the $CO$ and $cSBP$).

\newpage
To ease the optimization process, since the flow scales within arterial segments can greatly vary due to downstream branching, we can maintain similar scales within the domains by expressing the 1-D equations with respect to the cross-sectionally averaged velocity $u(x,t) = \frac{Q(x,t)}{A(x,t)}$ (derivation in the Appendix A). In reality, the human arterial network geometry is largely optimized such that the average blood velocity is maintained at a similar scale across different arterial segments, leading to uniform shear stress that lies within the physiological threshold. Thus the NS equations in terms of $u$ can be re-written as:

\begin{align}
\text{Momentum:}\quad
&\frac{\partial u}{\partial t}
+ \frac{u}{A}\,\frac{\partial A}{\partial t}
+ \frac{8}{3}u\,\frac{\partial u}{\partial x}
+ \frac{4}{3}\frac{u^2}{A}\,\frac{\partial A}{\partial x}
+ \frac{1}{\rho}\,\frac{\partial P}{\partial x}
+ \frac{8\mu}{\rho R^2}u
= 0, &
\label{p4:mom2}
\\[4pt]
\text{Continuity:}\quad
&\frac{\partial P}{\partial t}
+ \frac{1}{C_A(P)}\left(
\frac{\partial A^v}{\partial t}
+ \frac{\partial A}{\partial x}u
+ A\,\frac{\partial u}{\partial x}
\right)
= 0, &
\label{p4:cont2}
\\[4pt]
\text{Windkessel:}\quad
&\frac{\partial u}{\partial t}
+ \frac{u}{A^v}\,\frac{\partial A}{\partial t}
- \frac{1}{A R_1}\,\frac{\partial P}{\partial t}
- \frac{P}{A R_1 R_2 C_T}
- \left(1+\frac{R_1}{R_2}\right)\frac{u}{R_1 C_T}
= 0, &
\label{p4:wku}
\end{align}

However, since the scales of individual variables still have very different orders of magnitude within the cardiovascular beds, we can further enhance the neural network optimization by non-dimensionalizing the PDEs using dimensionless variables for space, time, velocity and pressure (see Appendix B). This technique has been previously used in PINNs to eliminate learning imbalances caused by dominant differential terms during the training process \cite{kissas2019machine}. Our neural network with parameters $\theta$ then aims to approximate the fields:
\[
u_\theta(x,t)\approx u(x,t),\qquad
P_\theta(x,t)\approx P(x,t),\qquad
A_\theta(x,t)\approx A(x,t).
\]
At interior collocation points $\{(x_j^{(m)},t_j^{(m)})\}_{j=1}^{N_{\mathrm{\mathcal{R}}}}$ for the momentum,
and $\{(x_j^{(c)},t_j^{(c)})\}_{j=1}^{N_{\mathrm{\mathcal{R}}}}$ for the continuity equation, we define the Momentum (Eq. \ref{p4:mom2}) and Continuity (Eq. \ref{p4:cont2}) residuals $\mathcal{R}_{\mathrm{mom}}, \mathcal{R}_{\mathrm{cont}}$, with the corresponding MSE loss expressions given by:

\begin{align}
\mathcal{L}_{\mathcal{R}}^{\mathrm{mom}}
&=
\frac{1}{N_{\mathrm{\mathcal{R}}}}
\sum_{j=1}^{N_{\mathrm{\mathcal{R}}}}
\left(
\mathcal{R}_{\mathrm{mom}}\big(x_j^{(m)},t_j^{(m)};\theta\big)
\right)^2,
\\[4pt]
\mathcal{L}_{\mathcal{R}}^{\mathrm{cont}}
&=
\frac{1}{N_{\mathrm{\mathcal{R}}}}
\sum_{j=1}^{N_{\mathrm{\mathcal{R}}}}
\left(
\mathcal{R}_{\mathrm{cont}}\big(x_j^{(c)},t_j^{(c)};\theta\big)
\right)^2,
\end{align}
and the total PDE residual loss of Eq. \ref{p4:eq:loss} is:
\begin{equation}
\mathcal{L}_{\mathcal{R}}
=
\mathcal{L}_{\mathcal{R}}^{\mathrm{mom}}
+
\mathcal{L}_{\mathcal{R}}^{\mathrm{cont}}.
\end{equation}

At each bifurcation node $k$, for times $\{t_j^{(k)}\}_{j=1}^{N_k}$, we also define mass and pressure conservation residuals as:

\begin{equation}
\mathcal{R}^{(k)}_{\mathrm{mass}}(t_j^{(k)};\theta)
=
A_{\theta,p}(L_p,t_j^{(k)})\,u_{\theta,p}(L_p,t_j^{(k)})
-
\sum_{i=1}^{N_d^{(k)}}
A_{\theta,d_i}(0,t_j^{(k)})\,u_{\theta,d_i}(0,t_j^{(k)}).
\end{equation}

\begin{equation}
\mathcal{R}^{(k,i)}_{\mathrm{press}}(t_j^{(k)};\theta)
=
P_{\theta,p}(L_p,t_j^{(k)})
-
P_{\theta,d_i}(0,t_j^{(k)}),
\qquad i=1,\dots,N_d^{(k)}.
\end{equation}

\noindent with their MSE contributions (summed over all junctions) given by:
\begin{align}
\mathcal{L}_{BC}^{\mathrm{mass}}
&=
\frac{1}{N_{BC}}
\sum_{k}
\sum_{j=1}^{N_k}
\left(
\mathcal{R}^{(k)}_{\mathrm{mass}}(t_j^{(k)};\theta)
\right)^2,
\\[4pt]
\mathcal{L}_{BC}^{\mathrm{press}}
&=
\frac{1}{N_{BC}}
\sum_{k}
\sum_{i=1}^{N_d^{(k)}}
\sum_{j=1}^{N_k}
\left(
\mathcal{R}^{(k,i)}_{\mathrm{press}}(t_j^{(k)};\theta)
\right)^2.
\end{align}

At each terminal site, the Windkessel condition \eqref{p4:wku} yields the following loss term:

\begin{equation}
\mathcal{L}_{BC}^{\mathrm{WK}}
=
\frac{1}{N_{\mathrm{WK}}}
\sum_{j=1}^{N_{\mathrm{WK}}}
\left(
\mathcal{R}_{\mathrm{WK}}(t_j;\theta)
\right)^2.
\end{equation}

Thus the total boundary-condition loss term of Eq. \ref{p4:eq:loss} is:
\begin{equation}
\mathcal{L}_{BC}
=
\mathcal{L}_{BC}^{\mathrm{mass}}
+
\mathcal{L}_{BC}^{\mathrm{press}}
+
\mathcal{L}_{BC}^{\mathrm{WK}}.
\end{equation}

For the data loss term of Eq. \ref{p4:eq:loss}, using cuff pressure measurements at a location $x_{\mathrm{cuff}}$, with systolic and diastolic targets $P_{\mathrm{SBP}}^{\mathrm{cuff}}$ and $P_{\mathrm{DBP}}^{\mathrm{cuff}}$, and associated times $t_{\mathrm{SBP}}$ and $t_{\mathrm{DBP}}$ in the simulated trace, we define:
\begin{equation}
\mathcal{L}_{data}
=
\frac{1}{2}
\left[
\left(
P_\theta(x_{\mathrm{cuff}})_{max}
- P_{\mathrm{SBP}}^{\mathrm{cuff}}
\right)^2
+
\left(
P_\theta(x_{\mathrm{cuff}})_{min}
- P_{\mathrm{DBP}}^{\mathrm{cuff}}
\right)^2
\right].
\end{equation}

Finally, since this work focuses on solving the inverse problem, we can't impose an exact initial condition at the inlet (i.e. cardiac flow pulse) but we can impose an arbitrary pulse shape which depends on the heart rate ($HR$) \cite{mertens1981influence} to ease the optimization process. Given an initial flow pulse $u(x,0) = \lambda_{u} \cdot u_{\mathrm{IC}}(x)$ expressed in terms of velocity, the parameter $\lambda_{u}$ is trainable along with the other network parameters, yielding the cardiac output ($CO$) which matches the patient-specific data (i.e. pressure data). Thus, on a set of initial-condition points $\{x_j^{(\mathrm{IC})}\}_{j=1}^{N_{\mathrm{IC}}}$, we impose:

\begin{equation}
\mathcal{L}_{IC}
=
\frac{1}{N_{\mathrm{IC}}}
\sum_{j=1}^{N_{\mathrm{IC}}}
\left(
u_\theta\big(x_j^{(\mathrm{IC})},0\big)
-
\lambda_{u} \cdot u_{\mathrm{IC}}\big(x_j^{(\mathrm{IC})}\big)
\right)^2.
\label{p4:flow}
\end{equation}

\subsection{Implementation details}

\subsubsection{Compliance - Pressure relationship}

To increase the efficiency of our PINN model, we observe that the constitutive law of Eq. \ref{p4:const} can be integrated to give an analytical expression relating $P$ and $A$ (derivation in the Appendix C):  

\begin{equation}
A(P)
=
A_{ref}
+
C_d
\left\{
a\,(P-P_0)
+
b\,P_w
\left[
\tan^{-1}\left(\frac{P-P_m}{P_w}\right)
-
\tan^{-1}\left(\frac{P_0-P_m}{P_w}\right)
\right]
\right\},
\end{equation}

\begin{equation}
\text{with} \quad C_d = \frac{A_{ref}}{\rho\,PWV^2}.
\end{equation}

By using this expression we can drop one neural network output for the area $A$, and calculate it explicitly during training based on the output for pressure $P$. This explicit coupling between $P$ and $A$ greatly enhances the stability of the optimization process as it directly constrains the pressure by the flexible expansion of the arteries, reducing any numerical instabilities from extreme pressures during the earlier training stages. 

\subsubsection{Parametrized PINNs as a neural operator}

To efficiently train a single network able to predict the flows within any of the arterial tree of Fig. \ref{p4:workflow}, we wish to train a neural operator which learns the family of solutions for all segments. In order to distinguish between different arterial vessels in the network, we employ a learnable embedding layer that maps discrete vessel indices to continuous representations. Thus, for a system with $N_v$ vessels (in this case $N_v = 8$), we define an embedding matrix:
\begin{equation}
\mathbf{E} \in \mathbb{R}^{N_v \times d_e},
\end{equation}

\noindent where $d_e$ is the embedding dimension (here $d_e = 1$). Each row $\mathbf{e}_i \in \mathbb{R}^{d_e}$ represents the learnable embedding vector for vessel $i$. For a given vessel index $v \in \{0, 1, \ldots, N_v-1\}$, the embedding operation is:
\begin{equation}
s = \mathbf{E}[v] = \mathbf{e}_v.
\end{equation}

This embedding vector $s$ is then concatenated with the temporal and spatial coordinates $(t, x)$ to form the input to the neural network:
\begin{equation}
\mathbf{h} = [t, x, s] \in \mathbb{R}^{3}
\end{equation}

The deep neural network $f_{\theta}$ then processes this augmented input to yield the predicted velocity and pressure fields:

\begin{equation}
 f_{\theta}(\mathbf{h}) = [u, p],
\end{equation}

The embedding matrix $\mathbf{E}$ is learned jointly with the network parameters during training, allowing the model to automatically discover vessel-specific features that distinguish hemodynamic characteristics across different arterial segments.

\subsubsection{Temporal periodicity}

\begin{figure}[H]
 \centering
 \includegraphics[width=0.55\textwidth]{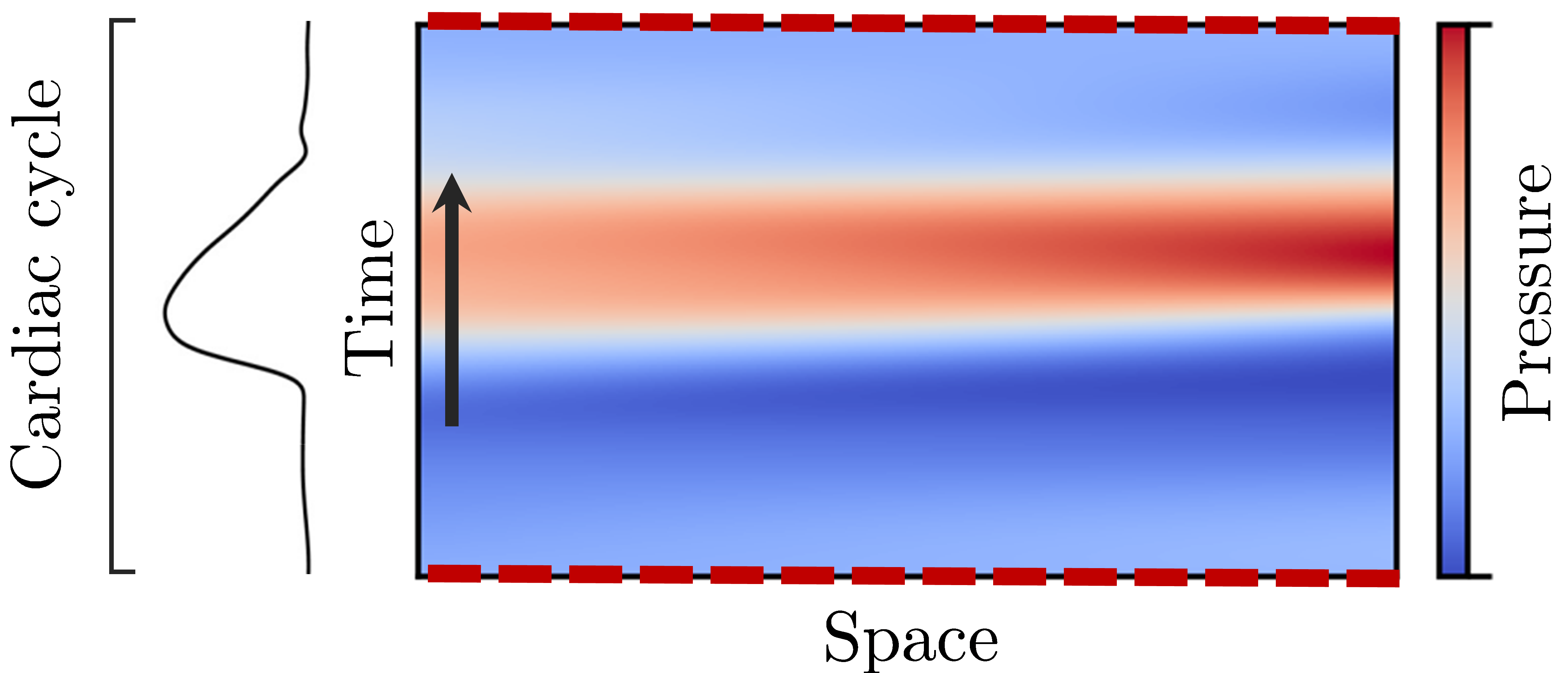}
 \caption[Temporal periodicity]{ \textbf{Temporal periodicity:} In PINNs we can enforce periodicity in the time dimension (red dashes) by transforming the input coordinates using Fourier feature embeddings. This achieves a periodic solution without the need of simulating multiple cardiac cycles as in traditional numerical integration.}
 \label{p4:temporal}
\end{figure}

One key difference of PINNs compared to traditional numerical methods, is that PINNs do not make a distinction between space and time, they simply treat both as different dimensions with equivalent properties. For example, during training, information can flow from the initial condition to a future time step (forward solution) in a similar fashion that it flows from an internal set of data points backward in time (inverse solution). On the other hand, during numerical integration, one must carefully implement an appropriate numerical scheme to move forward in time and converge to a stable solution. Although implicit schemes can guarantee stability, for periodic problems (e.g. heart flow), they require multiple cycles so that the algorithm settles in a solution which satisfies temporal periodicity. This is usually the most important criterion for the algorithm termination and is defined by ensuring that the flow and pressure at $t=0$ are equal to their corresponding values at $t=T$. On the other hand, with PINNs we can softly enforce periodicity by introducing a loss term that minimizes the residuals between two time steps, or enforce it exactly by using Fourier feature embeddings (Fig. \ref{p4:temporal}). The latter transforms the input using trigonometric functions which guarantee periodicity, and have been proven to be a significant enhancement for PINN performance \cite{dong2021method,anagnostopoulos2024residual}.

In this work, we apply Fourier feature embeddings to enforce temporal periodicity within every arterial segment. This way we are able converge to a solution for the entire arterial network within one cycle, which automatically satisfies the periodic constrains for $u$, $P$ and $A$ at the same time. The embeddings are given by: 

\begin{equation}
\mathbf{h} = \left[\cos(k_1 \omega t), \ldots, \cos(k_M \omega t), \sin(k_1 \omega t), \ldots, \sin(k_M \omega t), x, s\right]
\end{equation}

\noindent where $(t, x, s)$ are the input coordinates with $M$ Fourier modes, the harmonic numbers are $k_m = m$ for $m = 1, 2, \ldots, M$, and the fundamental angular frequency is:

\begin{equation}
\omega = \frac{2\pi}{T_{\text{cardiac}}}, \quad T_{\text{cardiac}} = \frac{N}{T_{\text{scale}}},
\end{equation}

\noindent with $N$ being the period parameter and $T_{\text{scale}}$ the temporal scaling factor.

The embedding transforms the input from dimension 3 to dimension $2M + 2$ before passing to the deep neural network. Our implementation uses $M=4$ modes, which yields a 10-dimensional input representation. The Fourier features explicitly provide periodic basis functions aligned with the cardiac cycle, enabling the network to overcome spectral bias and efficiently learn high-frequency temporal variations in pressure and flow waveforms. This effectively enhances the neural network's ability to capture oscillatory hemodynamic patterns and wave reflections.

\subsubsection{Patient-specific parameter estimation}

Performing an inverse solution for the identification of heart flow (Eq. \ref{p4:flow}) requires an adjusted patient-specific arterial network (geometry and compliance) through the known noninvasive measurements (age, height $cfPWV$), but also estimating the unknown patient parameters of the Windkessel condition (Eq. \ref{p4:wku}). For our model, these parameters are also learned during training, which increases the accuracy of the predictions. Thus, along with the learnable parameter of flow $\lambda_{u}$, the network learns two additional global scaling multipliers for $(C_T, R_T)$, namely $(\lambda_{C_T}, \lambda_{R_T})$. To constrain their range such that it lies within clinical experiments, we employ an arctan-based parameterization given by:
\begin{equation}
\lambda_{C_T} = c_{C_T} + h_{C_T}\frac{2}{\pi}\arctan(\theta_{C_T}), 
\qquad
\lambda_{R_T} = c_{R_T} + h_{R_T}\frac{2}{\pi}\arctan(\theta_{R_T}),
\end{equation}

\noindent where $(\theta_{C_T},\theta_{R_T})$ are unconstrained learnable parameters and
\begin{equation}
c = \frac{\lambda^{\min} + \lambda^{\max}}{2}, 
\qquad 
h = \frac{\lambda^{\max} - \lambda^{\min}}{2}.
\end{equation}

This transformation maps $\theta \in \mathbb{R}$ smoothly to the bounded interval 
$[\lambda^{\min},\lambda^{\max}]$ which is obtained from the literature:

\begin{equation}
\lambda_{C_T}^{\min} = 0.4, \quad \lambda_{C_{T0}}^{\max} = 2.265, \quad \lambda_{R_T}^{\min} = 0.2, \quad \lambda_{R_{T0}}^{\max} = 3.255
\end{equation}

These learnable parameters are therefore optimized jointly with the neural network weights to match clinical measurements ($CO$, cuff blood pressures), while the arctan transformation ensures physically admissible values throughout training.

\section{Results}

\subsection{PINN training}

For the optimization process, we train the neural network parameters, including the three learnable parameters $\lambda$, using a two-stage optimization: 100 iterations with Rprop and 4000 iterations with SSBroyden2. The latter has been proven to be a highly accurate BFGS version for PINN applications \cite{urban2025unveiling,kiyani2025optimizing}, hence it is adopted for the purposes of this study. The DNN architecture consists of 4 deep layers of 28 neurons each with 3 inputs, namely the position ($x$), time ($t$) and arterial segment ($s$). We note that similar performance was observed using the LBFGS algorithm, although using a larger architecture (5 deep layers of 128 neurons each). Thus, for the interest of memory we opted for the SSBroyden optimizer. As previously, mention the third input ($s$) enables the inference of the solution fields within different arterial segments using the same parametrized-DNN. 

\begin{figure}[H]
 \centering
 \includegraphics[width=0.5\textwidth]{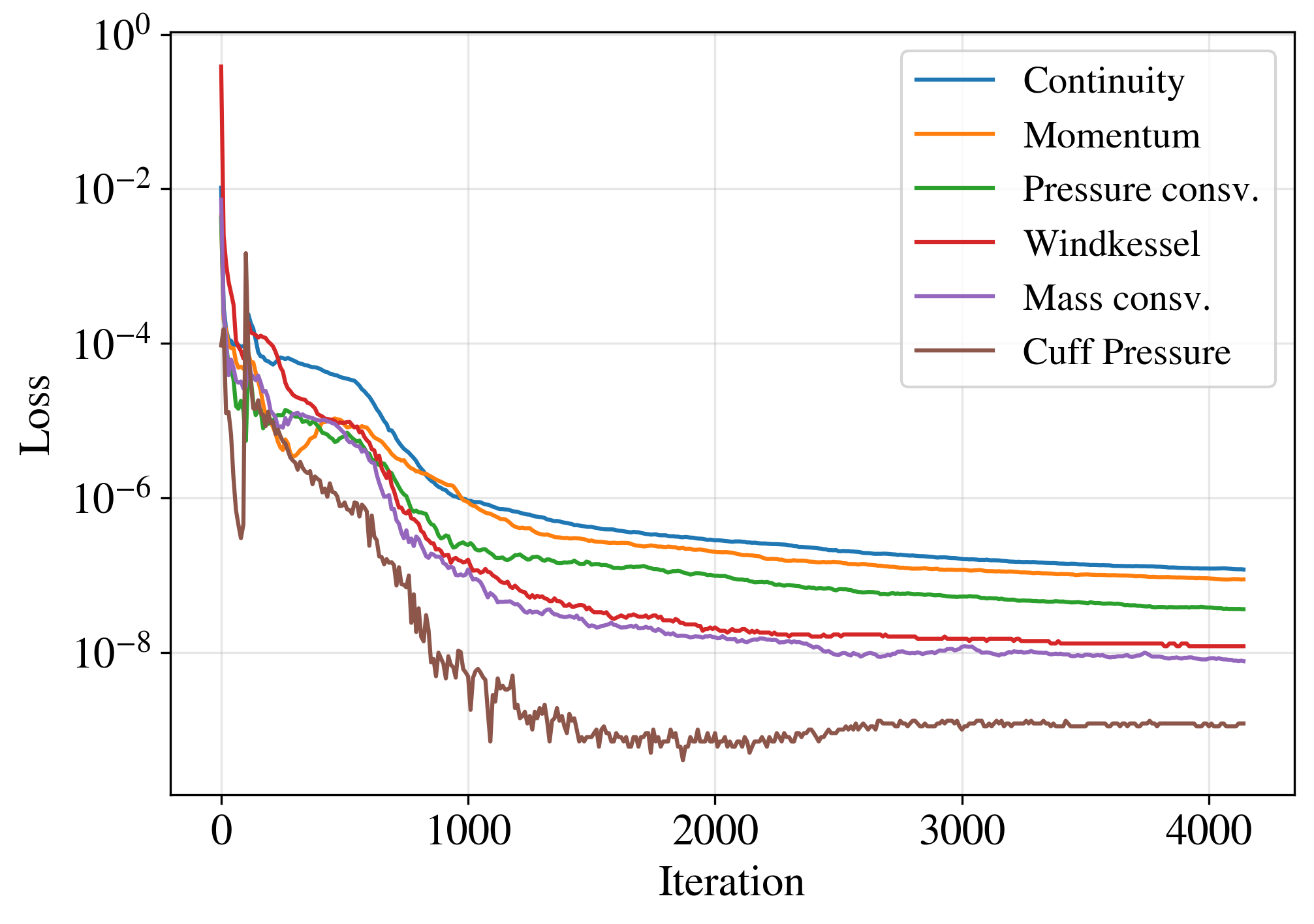}
 \caption[Training convergence]{ \textbf{Training convergence:} The full training takes roughly 4000 iterations: Rprop is used to warm up training for 100 iterations, followed by 4000 iterations of SSBroyden2. Each run takes approximately 5-7 minutes on a 4090 GPU.}
 \label{p4:convergence}
\end{figure}

The convergence plots of Fig. \ref{p4:convergence} indicate that Rprop can be used as an efficient alternative primer for LBFGS, which often struggles to converge during the early learning stages. While Adam is usually applied prior to $2^{nd}$ order optimizers, our experiments showed that it requires at least 10x times more iterations on average than Rprop, hence we opted for the latter. Moreover, the temporal periodic boundary condition enables the network to converge within just one heart cycle, as opposed to $\sim20$ cycles that the numerical solver usually takes. All loss terms appear to converge stably, indication that there are no major non-convexities in the loss landscape. Moreover, as the cuff pressure loss is essentially the only external information provided to the system, it rapidly decreases first and stabilizes while the rest of the equations are also minimized. The overall training time for obtaining a patient-specific solution is about 5-10 minutes on an Nvidia 4090 GPU. Notably, the computational time required is in the order of minutes compared to hours for a conventional inverse process using solely the 1-D solver and multiple of forward solutions \cite{bikia2019noninvasive,pagoulatou2018estimating} or compared to other state-of-the-art PINN implementations \cite{kissas2020machine,sarabian2022physics}.
 
\begin{figure}[H]
 \centering
 \includegraphics[width=0.75\textwidth]{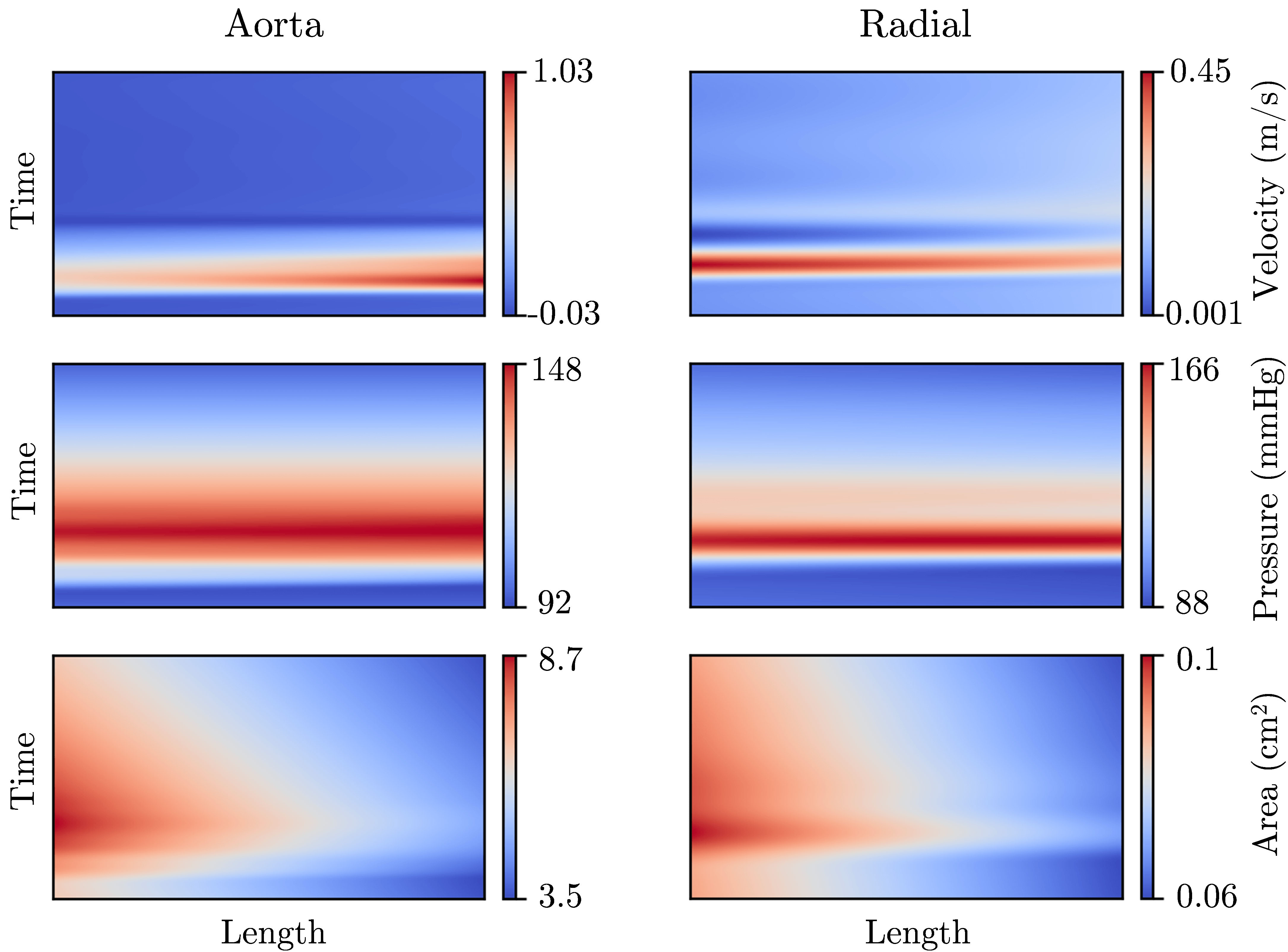}
 \caption[PINN solution fields]{ \textbf{PINN solution fields:} Indicative velocity ($u$), pressure ($P$) and the conical area ($A$) fields obtained by the PINN model for the aorta and radial arteries. Multiple wave reflections are visible mainly within the smaller radial artery.}
 \label{p4:indicative}
\end{figure}

In Figure \ref{p4:indicative}, we plot the indicative velocity ($u$), pressure ($P$) and area ($A$) fields for two of the eight domains (aorta and radial arteries), as solved by the PINN model. In our implementation, the truncated arterial tree has adjusted Windkessel parameters at its terminal sites to account for the missing branches from the complete arterial network. This should in theory provide all the necessary information of flow distribution within the major arteries to obtain an accurate solution of $CO$, even if a large portion of arterial segments (abdominal aorta, cerebral etc.) are omitted from the considered geometry.

\subsection{Numerical validation}

\begin{figure}[H]
 \centering
 \includegraphics[width=1.0\textwidth]{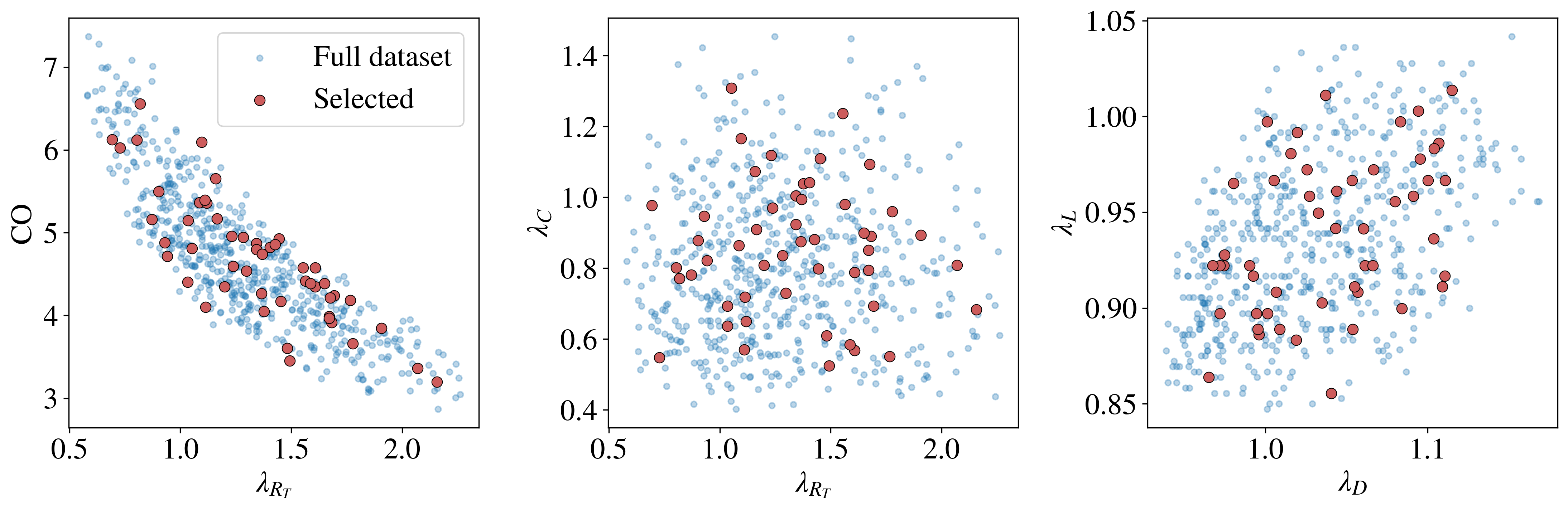}
 \caption[Numerical validation dataset]{ \textbf{Numerical validation dataset:} The initial \emph{in silico} dataset which was population-matched with the clinical Asklepios dataset, consisted of 620 patients. An LHS method was deployed to extract 50 representative patients for testing, showing good coverage for the indicative pair-wise parameter plots.}
 \label{p4:selection}
\end{figure}

To validate our inverse arterial model, we utilize an \emph{in silico} dataset, which has been generated with the input parameters drawn from the clinical population of Asklepios \cite{AsklepiosSegers2007}. This way we can test the validity of our implementation while also making sure the hemodynamic parameters of our virtual patients remain within physiological thresholds. Thus, starting from an initial dataset of 620 patients, we perform latin hypercube sampling (LHS) so that a highly representative subset of 50 patients is selected for the validation study, showing a good coverage for the indicative pair-wise parameter plots, as shown in Fig. \ref{p4:selection}.

\begin{figure}[H]
 \centering
 \includegraphics[width=0.725\textwidth]{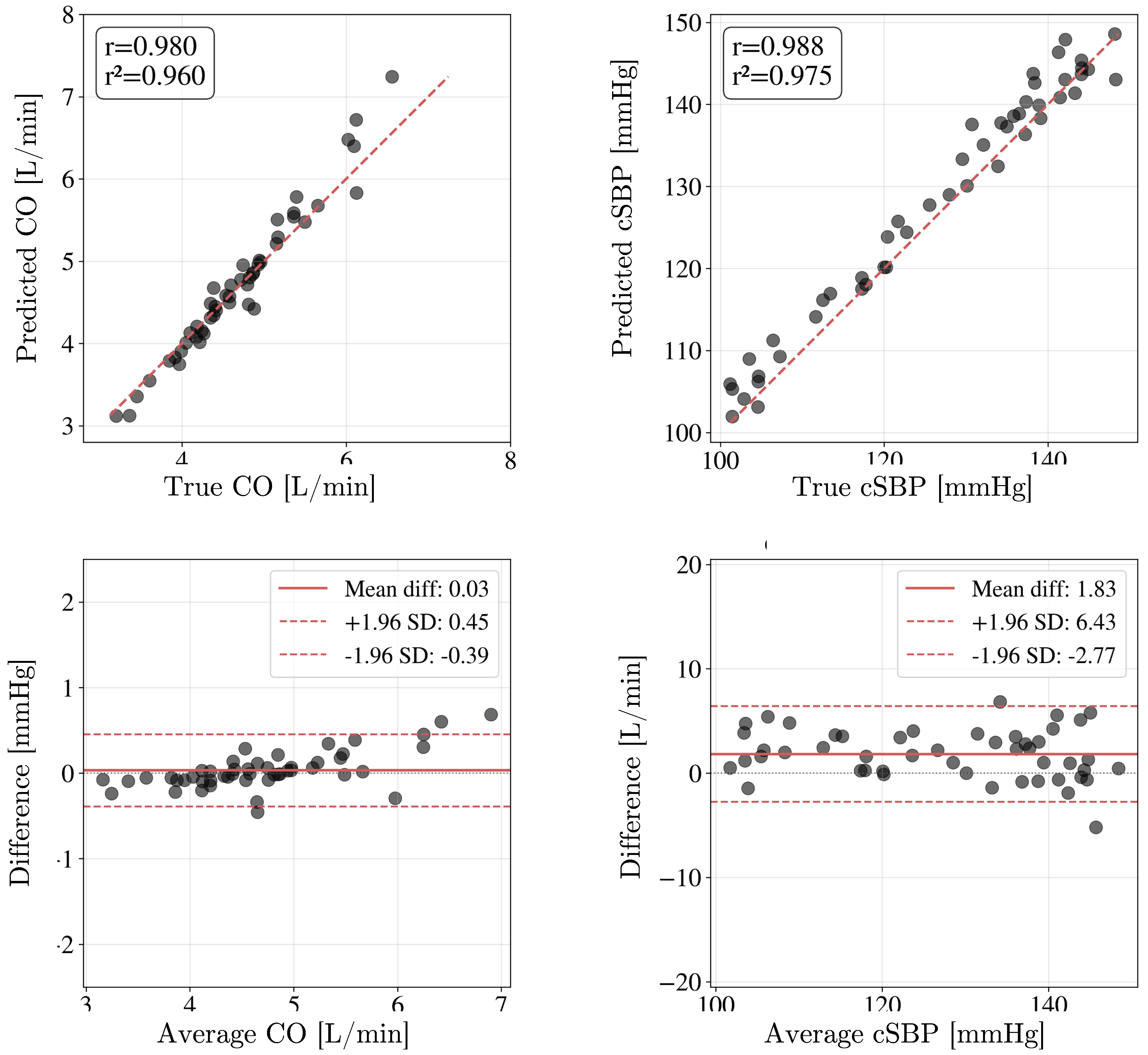}
 \caption[Numerical validation]{ \textbf{Numerical validation:} Pearson correlation and bland-altman plots for the estimation of $CO$ (left) and $cSBP$, showing a near perfect agreement, with the expection of a few high $CO$ values and a minor systematic underestimation of the $cSBP$, which could be attributed to the truncated geometry approximation.}
 \label{p4:numerical}
\end{figure}

To enable a fair comparison, we perform the steps of Fig. \ref{p4:workflow} by adjusting each patient-specific characteristics and solving for $CO$, given the $(DBP, SBP)$ values at the brachial artery, while keeping the Windkessel parameters fixed. Overall, the inverse model shows near-perfect correlation between the predictions (Fig. \ref{p4:numerical}) and the true numerical values. The Pearson correlation of 0.981 for $CO$ and 0.988 for $cSBP$ validates our PINN implementation, including all the assumptions and derivations made in the process. We note that the the slight underestimation of the $cSBP$ and $CO$ (only for higher values) is likely attributed to the truncated geometry approximation.

\subsection{Clinical study}

\begin{figure}[H]
 \centering
 \includegraphics[width=0.725\textwidth]{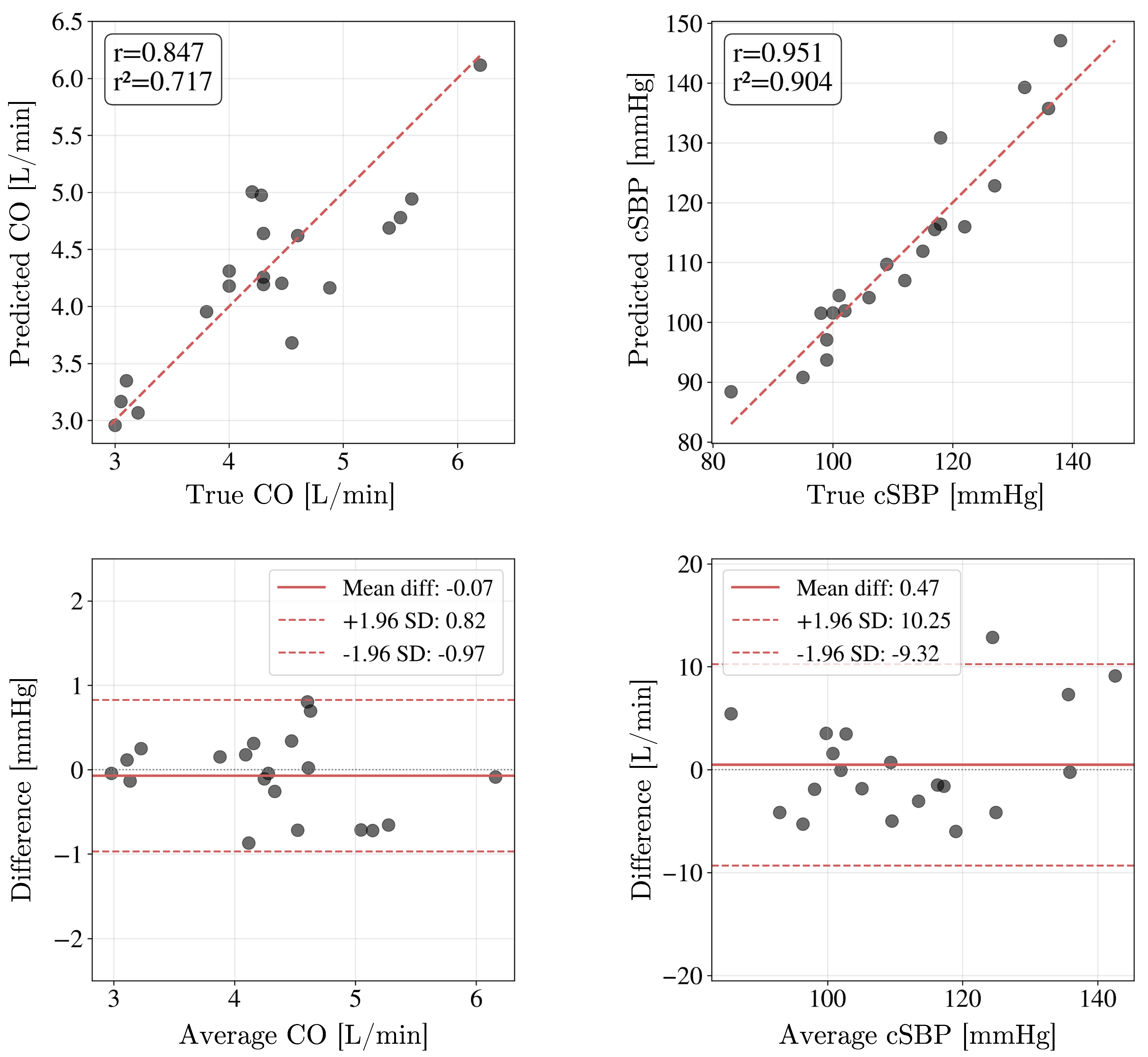}
 \caption[Clinical validation]{ \textbf{Clinical validation:} Pearson correlation and bland-altman plots for the estimation of $CO$ (left) and $cSBP$, showing a good agreement between the predictions and the true values. The errors of the $CO$ could be partially attributed to the truncated geometry approximation, which does not perfectly follow the true patient-specific scaling of the clinical dataset.}
 \label{p4:clinical}
\end{figure}

The accurate estimation of central cardiovascular parameters such as $CO$ and $cSBP$ is of high clinical relevance because they directly reflect cardiovascular health and the true hemodynamic load imposed on the heart and vital organs. Central blood pressure, unlike peripheral measurements, more accurately captures the effects of arterial stiffness and wave reflections, and has been shown to be more strongly associated with cardiovascular events and target-organ damage (e.g., left ventricular hypertrophy and renal dysfunction) \cite{nichols2022mcdonald,safar2003current}. Cardiac output is a fundamental determinant of tissue perfusion and is central to diagnosis, risk stratification, and therapeutic decision-making in conditions such as heart failure, shock, and perioperative critical illness \cite{cecconi2014consensus}. In this context, inverse blood flow modeling that estimates central quantities from peripheral measurements is clinically appealing, as it enables access to physiologically meaningful parameters with reduced invasiveness. In Fig. \ref{p4:clinical} we show a comparison between our model predictions and the clinical dataset obtained from a previous study \cite{papaioannou2014first}.

To enhance the predictions on the clinical dataset, the aortic stiffening was modeled with stiffness increase of the proximal aorta \cite{kaess2012aortic}. Prior work has shown that this age-related stiffness heterogeneity significantly influences central pressure dynamics and arterial wave reflections \cite{reymond2012systolic}. To capture this behavior, age-dependent measurements of regional aortic stiffening were adopted from \cite{kimoto2003preferential}. The heterogeneous stiffening pattern was implemented by adjusting the local distensibility of the proximal aortic segments using an age-specific proximal scaling factor which multiplied all the aortic segments, on top of the global factor that is applied uniformly across the entire arterial tree.

Overall, the predictions achieve a satisfactory correlation for the central characteristics of both $CO$ and $cSBP$ ($r=0.847, r=0.951$, respectively). Solving the inverse problem offers an advantage for estimating the $CO$ when the terminal resistance is not known, since the full flow field is solved to match the patient-specific measurements. This suggests that essential information which is unavailable to a surrogate model, can only be recovered by enforcing and minimizing the PDE residuals. Finally, unlike surrogate models, the PINN framework does not rely on clinically pooled data; instead, it directly infers patient-specific hemodynamics that best fit the measurements. This is particularly advantageous when population data are scarce or misaligned with the study objectives. Moreover, disease or age-specific effects can be incorporated directly into the numerical model itself, rather than indirectly through the generation of a virtual training dataset.

\section{Strengths and limitations}

PINNs offer several strengths for cardiovascular modeling and related inverse problems. A key advantage is noninvasiveness: they can combine routine, noninvasive measurements (e.g., MRI-derived flow, cuff pressure) with governing equations, reducing reliance on catheterization or other invasive procedures. That same physics constraint can regularize learning, so predictions can remain robust even when the available data are sparse or noisy. PINNs can also be efficient relative to purely data-driven ML because the model ``borrows” information from the physical laws, often reducing the amount of patient data required. In practice, this can make patient-specific parameter inference (e.g., cardiac output, arterial compliance, peripheral resistance) feasible within a single training run on the order of minutes, whereas classical inverse parameter fitting with repeated numerical simulations can be substantially more time-consuming. Beyond parameter identification, PINNs can act as a bridge between fidelity levels by injecting high-fidelity information (e.g., 4D MRI or CFD) to correct or augment lower-fidelity solvers such as 1D Navier-Stokes models, and they can be updated as physiology changes (e.g., stiffening or disease progression), supporting more credible longitudinal predictions.

The main limitations are tied to optimization stability and the amount of problem-specific fine-tuning required. While the basic formulation is straightforward for simple geometries and well-posed boundary/initial conditions, performance can degrade sharply for complex settings involving multi-domain coupling, heterogeneous boundary conditions, stiff dynamics, or very limited observations. In these regimes, training can become sensitive to loss weighting, sampling strategies, network architecture, and the particular choice of ``enhancements” (e.g., adaptive weights, domain decomposition, optimizers), and there is no universal configuration that guarantees stability or improved accuracy. As a result, significant effort often goes into empirical testing of hyperparameters and stabilization tricks, to avoid brittle convergence, with solutions that can be inconsistent across runs or fail to capture sharp features unless carefully engineered.

\section{Summary}

This work shows that central hemodynamic quantities can be inferred noninvasively and in a patient-specific manner by coupling a validated 1-D arterial network model with physics-informed neural networks. By embedding the governing 1-D flow and pressure equations, the bifurcation conservation laws, and the Windkessel terminal behavior directly into the learning objective, the proposed inverse PINN reconstructs pressure and flow fields across an arterial tree using minimal peripheral information from a cuff measurement. The framework therefore provides a competitive alternative to population-averaged transfer-function reconstructions and reduces the dependence on invasive catheter-based assessments, while remaining consistent with physiological constraints.

Aside from accuracy, the main contribution is practicality for personalized inference. The method solves the full eight-artery truncated network with a single neural model and learns key patient-specific quantities, like the aortic $CO$ and $SBP$, alongside global multipliers for terminal resistance and compliance, within a single training run. With the proposed stabilization and efficiency enhancements, convergence is achieved in roughly 4000 iterations, delivering solutions in minutes and offering an order-of-magnitude speedup over iterative inverse approaches that rely on repeated forward 1-D simulations. These results support the feasibility of near real-time, noninvasive hemodynamic assessment and motivate future validation on larger and more diverse clinical datasets, and deployment scenarios such as longitudinal monitoring and wearable-device decision support. Future versions of this work will aim at incorporating clinically measured pulse pressure waves at the radial artery, in order to couple numerical simulations with live patient-specific health monitoring.

\section*{Acknowledgements}
We thank Profs. T. G. Papaioannou and A. D. Protogerou from the Medical School of the National and Kapodistrian University of Athens, for providing the cardiac output patient dataset of the clinical study. 

\section*{Funding}
This work has received funding from the Swiss State Secretariat for Education, Research and Innovation (SERI), in the context of the EU-funded VITAL project.

\bibliographystyle{elsarticle-num} 
\bibliography{cas-refs}

% \newpage
\appendix
% \setcounter{section}{0}
% \appendix
% \newpage
\section{Dimensional equations in terms of u(x,t)}

Starting from
\[
Q(x,t) = A(x,t)\,u(x,t),
\]
and the 1-D momentum and continuity equations (neglecting gravity for simplicity):

\begin{equation}
\frac{\partial Q}{\partial t}
+
\frac{\partial}{\partial x}\left( \int_A u^2\,\mathrm{d}A \right)
=
-\frac{A}{\rho}\,\frac{\partial P}{\partial x}
- 2\pi R\,\frac{\mu}{\rho}\,\frac{\partial u}{\partial r}\Bigg|_{r=R},
\label{p4:eq:dim_Q_mom}
\end{equation}
\begin{equation}
\frac{\partial P}{\partial t}
=
-\frac{1}{C_A}\left(
\frac{\partial A^v}{\partial t}
+
\frac{\partial Q}{\partial x}
\right),
\label{p4:eq:dim_P_cont}
\end{equation}
and the Windkessel equation at a terminal site
\begin{equation}
\frac{\partial Q_T}{\partial t}
=
\frac{1}{R_1}\,\frac{\partial P_T}{\partial t}
+
\frac{P_T}{R_1 R_2 C_T}
+
\left(1+\frac{R_1}{R_2}\right)\frac{Q_T}{R_1 C_T}.
\label{p4:eq:dim_WK_Q}
\end{equation}

\medskip
Assuming a fully-developed axisymmetric parabolic velocity profile,
\[
\int_A u^2\,\mathrm{d}A = \frac{4}{3}A\,u^2,
\qquad
\frac{\partial u}{\partial r}\Big|_{r=R} = -\frac{4u}{R},
\]
we write all equations in terms of $u$.

\smallskip
\textbf{Momentum in terms of $u$.} Since $Q=A u$, \eqref{p4:eq:dim_Q_mom} becomes
\[
\frac{\partial Q}{\partial t} = A\,\frac{\partial u}{\partial t} + u\,\frac{\partial A}{\partial t},
\qquad
\frac{\partial}{\partial x}\left(\int_A u^2\,\mathrm{d}A \right)
=
\frac{\partial}{\partial x}\left(\frac{4}{3}A u^2\right)
=
\frac{4}{3}\left(
\frac{\partial A}{\partial x}u^2 + 2A u\,\frac{\partial u}{\partial x}
\right).
\]
We divide by $A$ and use $A = \pi R^2$:
\begin{equation}
\frac{\partial u}{\partial t}
+ \frac{u}{A}\,\frac{\partial A}{\partial t}
+ \frac{8}{3}u\,\frac{\partial u}{\partial x}
+ \frac{4}{3}\frac{u^2}{A}\,\frac{\partial A}{\partial x}
+ \frac{1}{\rho}\,\frac{\partial P}{\partial x}
+ \frac{8\mu}{\rho R^2}u
= 0.
\label{p4:eq:dim_u_mom}
\end{equation}

\smallskip
\textbf{Continuity/constitutive in terms of $u$.} Using $Q = A u$ in \eqref{p4:eq:dim_P_cont}:
\[
\frac{\partial Q}{\partial x}
=
\frac{\partial}{\partial x}(A u)
=
\frac{\partial A}{\partial x}u + A\,\frac{\partial u}{\partial x},
\]
so
\begin{equation}
\frac{\partial P}{\partial t}
+
\frac{1}{C_A}\left(
\frac{\partial A}{\partial t}
+
\frac{\partial A}{\partial x}u
+
A\,\frac{\partial u}{\partial x}
\right)
= 0.
\label{p4:eq:dim_u_cont}
\end{equation}

\smallskip
\textbf{Windkessel in terms of $u$.} At a terminal section, $Q_T = A_T u_T$ gives
\[
\frac{\partial Q_T}{\partial t}
=
A_T\,\frac{\partial u_T}{\partial t}
+
u_T\,\frac{\partial A_T}{\partial t}.
\]
Substituting into \eqref{p4:eq:dim_WK_Q} and dropping the subscript $T$:
\begin{equation}
A\,\frac{\partial u}{\partial t}
+
u\,\frac{\partial A}{\partial t}
=
\frac{1}{R_1}\,\frac{\partial P}{\partial t}
+
\frac{P}{R_1 R_2 C_T}
+
\left(1+\frac{R_1}{R_2}\right)\frac{A u}{R_1 C_T},
\end{equation}
which can be written as
\begin{equation}
\frac{\partial u}{\partial t}
+
\frac{u}{A}\,\frac{\partial A}{\partial t}
-
\frac{1}{A R_1}\,\frac{\partial P}{\partial t}
-
\frac{P}{A R_1 R_2 C_T}
-
\left(1+\frac{R_1}{R_2}\right)\frac{u}{R_1 C_T}
= 0.
\label{p4:eq:dim_u_WK}
\end{equation}

\smallskip
\textbf{Conservation of mass and pressure at junctions.} At a bifurcation node (parent $p$, daughters $d_i$):
\begin{equation}
A_p(L_p,t)\,u_p(L_p,t)
=
\sum_{i=1}^{N_d} A_{d_i}(0,t)\,u_{d_i}(0,t),
\label{p4:eq:dim_mass_junc}
\end{equation}
\begin{equation}
P_p(L_p,t) = P_{d_i}(0,t),\qquad i=1,\dots,N_d.
\label{p4:eq:dim_press_junc}
\end{equation}

\section{Non-dimensionalization}

We now introduce dimensionless variables for space, time, velocity and pressure:

\[
x = L\,x^*,\qquad
t = \frac{L}{U}\,t^*,\qquad
u = U\,u^*,\qquad
P = \rho U^2\,P^*.
\]

Derivatives transform as
\[
\frac{\partial}{\partial t}
= \frac{U}{L}\,\frac{\partial}{\partial t^*},
\qquad
\frac{\partial}{\partial x}
= \frac{1}{L}\,\frac{\partial}{\partial x^*},
\]
so
\[
\frac{\partial u}{\partial t}
= \frac{U^2}{L}\,\frac{\partial u^*}{\partial t^*},
\quad
\frac{\partial u}{\partial x}
= \frac{U}{L}\,\frac{\partial u^*}{\partial x^*},
\quad
\frac{\partial P}{\partial t}
= \frac{\rho U^3}{L}\,\frac{\partial P^*}{\partial t^*},
\quad
\frac{\partial P}{\partial x}
= \frac{\rho U^2}{L}\,\frac{\partial P^*}{\partial x^*},
\]
while
\[
\frac{\partial A}{\partial t}
= \frac{U}{L}\,\frac{\partial A}{\partial t^*},
\qquad
\frac{\partial A}{\partial x}
= \frac{1}{L}\,\frac{\partial A}{\partial x^*},
\]

\medskip
\textbf{Non-dimensional momentum equation.}

Substitute into \eqref{p4:eq:dim_u_mom}:
\begin{align*}
&\frac{U^2}{L}\frac{\partial u^*}{\partial t^*}
+ \frac{U u^*}{A}\,\frac{U}{L}\frac{\partial A}{\partial t^*}
+ \frac{8}{3}U u^*\,\frac{U}{L}\frac{\partial u^*}{\partial x^*}
+ \frac{4}{3}\frac{U^2 u^{*2}}{A}\,\frac{1}{L}\frac{\partial A}{\partial x^*}
\\
&\quad
+ \frac{1}{\rho}\,\frac{\rho U^2}{L}\frac{\partial P^*}{\partial x^*}
+ \frac{8\mu}{\rho R^2}U u^*
= 0.
\end{align*}
We factor $U^2/L$ from the first five terms and divide the whole equation by $U^2/L$:
\begin{equation}
\frac{\partial u^*}{\partial t^*}
+
\frac{u^*}{A}\,\frac{\partial A}{\partial t^*}
+
\frac{8}{3}u^*\,\frac{\partial u^*}{\partial x^*}
+
\frac{4}{3}\frac{u^{*2}}{A}\,\frac{\partial A}{\partial x^*}
+
\frac{\partial P^*}{\partial x^*}
+
\frac{8\mu L}{\rho U R^2}\,u^*
= 0.
\label{p4:eq:nd_u_mom}
\end{equation}

\medskip
\textbf{Non-dimensional continuity equation.}

From \eqref{p4:eq:dim_u_cont}:
\[
\frac{\partial P}{\partial t}
+
\frac{1}{C_A}\left(
\frac{\partial A^v}{\partial t}
+
\frac{\partial A}{\partial x}u
+
A\,\frac{\partial u}{\partial x}
\right)
= 0.
\]
Substituting the scalings:
\[
\frac{\rho U^3}{L}\frac{\partial P^*}{\partial t^*}
+
\frac{1}{C_A}\left[
\frac{U}{L}\frac{\partial A^v}{\partial t^*}
+
\frac{1}{L}\frac{\partial A}{\partial x^*}(U u^*)
+
A\,\frac{U}{L}\frac{\partial u^*}{\partial x^*}
\right]
= 0.
\]
Multiplying by $L/(\rho U^3)$:
\begin{equation}
\frac{\partial P^*}{\partial t^*}
+
\frac{1}{\rho U^2 C_A}\left(
\frac{\partial A^v}{\partial t^*}
+
\frac{\partial A}{\partial x^*}u^*
+
A\,\frac{\partial u^*}{\partial x^*}
\right)
= 0.
\label{p4:eq:nd_u_cont_full}
\end{equation}

\medskip
\textbf{Non-dimensional Windkessel equation.}

Starting from \eqref{p4:eq:dim_u_WK}:
\[
\frac{\partial u}{\partial t}
+
\frac{u}{A}\,\frac{\partial A}{\partial t}
-
\frac{1}{A R_1}\,\frac{\partial P}{\partial t}
-
\frac{P}{A R_1 R_2 C_T}
-
\left(1+\frac{R_1}{R_2}\right)\frac{u}{R_1 C_T}
= 0.
\]
Substituting the scalings:
\[
\frac{U^2}{L}\frac{\partial u^*}{\partial t^*}
+
\frac{U u^*}{A}\,\frac{U}{L}\frac{\partial A}{\partial t^*}
-
\frac{1}{A R_1}\left(\frac{\rho U^3}{L}\frac{\partial P^*}{\partial t^*}\right)
-
\frac{\rho U^2 P^*}{A R_1 R_2 C_T}
-
\left(1+\frac{R_1}{R_2}\right)\frac{U u^*}{R_1 C_T}
= 0.
\]
Multiplying by $L/U^2$:
\begin{equation}
\frac{\partial u^*}{\partial t^*}
+
\frac{u^*}{A}\,\frac{\partial A}{\partial t^*}
-
\frac{\rho U}{A R_1}\,\frac{\partial P^*}{\partial t^*}
-
\frac{\rho L}{A R_1 R_2 C_T}\,P^*
-
\left(1+\frac{R_1}{R_2}\right)\frac{L}{U R_1 C_T}\,u^*
= 0.
\label{p4:eq:nd_u_WK}
\end{equation}

This is the non-dimensional Windkessel condition in terms of $u^*$, $P^*$, and the \emph{dimensional} $A$, $R_1$, $R_2$, $C_T$.

\medskip
\textbf{Conservation laws at junctions.}

From \eqref{p4:eq:dim_mass_junc}-\eqref{p4:eq:dim_press_junc}, using $Q=A u$ and the same scalings (note that common factors cancel), the junction conditions retain their form:
\begin{equation}
A_p(L_p,t)\,u_p(L_p,t)
=
\sum_{i=1}^{N_d} A_{d_i}(0,t)\,u_{d_i}(0,t),
\qquad
P_p(L_p,t) = P_{d_i}(0,t),
\end{equation}
with $x$ and $t$ denoting the scaled variables $(x^*,t^*)$.
\section{Analytical constitutive law}

We start from the constitutive law where $A$ in the RHS has been replaced by
$A_{ref}$:
\begin{equation}
\frac{\mathrm{d}A}{\mathrm{d}P}
=
\frac{A_{ref}}{\rho\,PWV^2}
\left[
a + \frac{b}{1+\left(\dfrac{P-P_m}{P_w}\right)^2}
\right],
\label{p4:eq:CA_mod}
\end{equation}
with boundary condition
\begin{equation}
A(P_0) = A_{ref}.
\label{p4:eq:BC}
\end{equation}

We define
\[
C_d := \frac{A_{ref}}{\rho\,PWV^2}.
\]

and integrate \eqref{p4:eq:CA_mod} from $P_0$ to $P$:
\begin{equation}
A(P) - A_{ref}
=
\int_{P_0}^{P} \frac{\mathrm{d}A}{\mathrm{d}\tilde P}\,\mathrm{d}\tilde P
=
C_d \int_{P_0}^{P}
\left[
a + \frac{b}{1+\left(\dfrac{\tilde P-P_m}{P_w}\right)^2}
\right]
\mathrm{d}\tilde P.
\end{equation}

We split the integral:
\begin{align}
A(P) - A_{ref}
&=
C_d \left\{
a \int_{P_0}^{P} \mathrm{d}\tilde P
+
b \int_{P_0}^{P}
\frac{1}{1+\left(\dfrac{\tilde P-P_m}{P_w}\right)^2}
\mathrm{d}\tilde P
\right\} \\
&=
C_d \left\{
a(P-P_0)
+
b P_w
\left[
\tan^{-1}\left(\frac{P-P_m}{P_w}\right)
-
\tan^{-1}\left(\frac{P_0-P_m}{P_w}\right)
\right]
\right\}.
\end{align}

Thus
\begin{equation}
A(P)
=
A_{ref}
+
C_d
\left\{
a(P-P_0)
+
b P_w
\left[
\tan^{-1}\left(\frac{P-P_m}{P_w}\right)
-
\tan^{-1}\left(\frac{P_0-P_m}{P_w}\right)
\right]
\right\}.
\label{p4:eq:A_final_general}
\end{equation}

For the specific coefficients used in this study $a=0.4$, $b=5$, this becomes:

\begin{equation}
A(P)
=
A_{ref}
+
C_d
\left\{
0.4\,(P-P_0)
+
5\,P_w
\left[
\tan^{-1}\left(\frac{P-P_m}{P_w}\right)
-
\tan^{-1}\left(\frac{P_0-P_m}{P_w}\right)
\right]
\right\},
\end{equation}

% \begin{equation}
% \text{with} \quad C_d = \frac{A_{ref}}{\rho\,PWV^2}.
% \end{equation}

% \end{appendices}

% \newpage
% {\fontsize{12}{12}\selectfont
%     \setlength{\bibsep}{0pt} 
%   \bibliographystyle{elsarticle-num}
%   \bibliography{cas-refs}
% }
 
\end{document}